\documentclass[onecolumn,showpacs,preprintnumbers,amsmath,amssymb]{revtex4}
\usepackage [dvips]{graphicx}

\usepackage [english]{babel}

\begin{document}

\title{Effect of dielectric confinement on energetics of quantum metal films}

\author{A. V. Babich, P. V. Vakula, and V. V. Pogosov\footnote{Corresponding author: E-mail:
 vpogosov@zstu.edu.ua }}

\address{Department of Micro- and Nanoelectronics, Zaporozhye National
Technical University, Zhukovsky Str. 64, Zaporozhye 69063, Ukraine}

\date{\today}

\begin{abstract}

We suggest a method for the self-consistent calculations of
characteristics of metal films in dielectric environment. Within a
modified Kohn-Sham method and stabilized jellium model, the most
interesting case of asymmetric metal-dielectric sandwiches is
considered, for which dielectric media are different from the two
sides of the film. As an example, we focus on Na, Al and Pb. We
calculate the spectrum, electron work function, and surface energy
of polycrystalline and crystalline films placed into passive
isolators. We find that a dielectric environment generally leads to
the decrease of both the electron work function and surface energy.
It is revealed that the change of the work function is determined
only by the average of dielectric constants from both sides of the
film.


\end{abstract}

\maketitle

\section{INTRODUCTION}

Thin metal films and flat islands on semiconductor or dielectric
substrates can be considered as two-dimensional electron systems
with properties, which are of interest both from the fundamental
point of view and from the perspective of their application in
nanoscale electronic devices.

There are a limited number of experimental works focused on quantum
size effects in such systems (for reviews, see
\cite{Otero,14.,BUUUU,LLLLLLL,TAAAA,VAAAA,Viiii,CHHHH,Chuu}) due to
difficulties in sample fabrication, as well as because of lack of
suitable experimental methods. One of the most important
characteristics of metal nanostructures is electron work function.

There are different methods, which enable one to calculate electron
structure of slabs consisting of few monoatomic layers. Let us
combine them into three groups according to the complexity of
computations: I -- the Sommerfeld electrons in-a-box model
(analytical calculations, slabs and wires)
\cite{1.,2.,PKV,Han,KuPo,ddddddd}; II -- self-consistent
calculations within various versions of jellium model (slabs and
wires) \cite{Schulte,11.,Fiolhais2000,Smog,Horowitz2009}; III --
\emph{ab initio} calculations (slabs) \cite{8.,9.,ZHHHH,10.}. The
obtained results are illustrated in figure 1 for all these three
groups. An important ingredient of approaches within group III is
the monolayer number in the film (see dots in figure 1). For groups
I and II, $L$ changes continuously.

In group I, the Fermi energy (kinetic energy) $\varepsilon_{\rm
F}(L)$ is counted from the \emph{flat} bottom of conductivity band,
while the work function $W(L)$ is counted from the vacuum level.
Therefore, their size dependencies are ``asymmetric''. In addition
to quantum oscillations, these quantities contain monotonic size
contributions, which, at small film thicknesses, \emph{together}
show up through inequalities $0<W(L)<W_{0}$ and $\varepsilon_{\rm
F}(L)>\bar{\varepsilon}_{\rm F}>0$, where $W_{0}$ and
$\bar{\varepsilon}_{\rm F}$ correspond to the three dimensional (3D)
metal (allowing for the energy counting for $\bar{\varepsilon}_{\rm
F}$).

In  \cite{pppppp,1996}, an asymptotic behavior of electron chemical
potential for spherical clusters of radius $R$ was determined, from
which it follows that
\begin{equation}
W(R)=W_{0}- \frac{c_{1}}{R}<W_{0}, \label{basic}
\end{equation}
where  $c_{1}\simeq 2.5\,{\rm eV}\times a_{0}$ for simple metals,
$a_{0}=\frac{\hbar^{2}}{me^{2}}$. It is expectable that such a
monotonic contribution must appear for films also. However, in
contrast to the case of group I, self-consistent calculations of
groups II and III (see figure 1), at small film thickness, point out
to the suppression of monotonic dependence (having an asymptotic
(\ref{basic})) by corrections of higher orders of smallness. For
instance, compensation of terms
$-\frac{c_{1}}{L}+\frac{c_{2}}{L^{2}}$ occurs at
$L^{*}=\frac{c_{2}}{c_{1}}$, and $L^{*}$ is large, provided
$c_{2}\gg c_{1}>0$.

Experimental results also do not allow to draw unambiguous
conclusions on the character of monotonic component of $W(L)$: in
experiments  \cite{BUUUU}, it is absent (Yb films on Si substrate),
while, according to \cite{14.,TAAAA}, it coincides with the one of
group I. Note that the comparison of a measured work function for
the sandwich consisting of Ag film on Fe(100) in \cite{14.,TAAAA}
with calculated results for slabs \emph{in vacuum} is rather
relative.

Work function as well as film surface energy in dielectric
environment have never been calculated before. The aim of this work
is to compute energy characteristics of metal films in dielectrics.
We suggest a method for self-consistent calculations of equilibrium
profiles of electron concentration, effective potential, energy
spectrum, and integral characteristics of metal films in dielectrics
and dielectric substrates. The developed method is based on a
stabilized jellium model \cite{PTS} and local density approximation
for exchange-correlation potential \cite{Per-Zug}, which were used
by us before \cite{Bab-2008} to analyze characteristics of
semi-infinite metal with dielectric coating.

This paper is organized as follows. In Section II, we formulate our
model. In Section III, we presents our main results and provide a
discussion of them. We conclude in Section IV.

%%%%%%%%%%%%%%%%%%%%%%%%%%%%%%%%%%%%%%%%%%%%%%%%%%%%%%%%%%
\begin{figure}[!t!b!p]
\centering
\includegraphics [width=.4\textwidth]{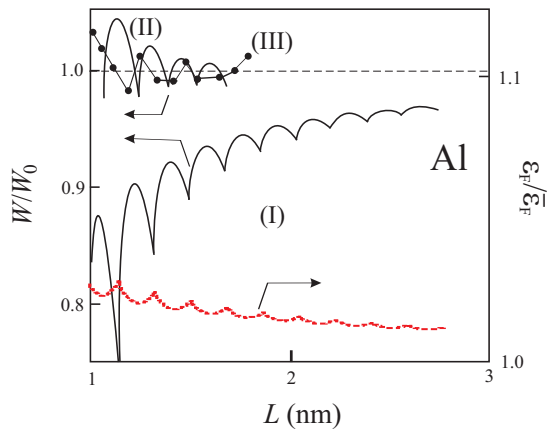}
\caption{Illustration of the computation results for groups I, II
and III (data for group I are deduced from  \cite{PKV}).}
\label{Fig1}
\end{figure}
%%%%%%%%%%%%%%%%%%%%%%%%%%%%%%%%%%%%%%%%%%%%%%%%%%%%%%%%%%

\section{MODEL}

Let us consider a metallic film of thickness $L$ at zero
temperature. We direct $z-$axis perpendicularly to the film surface
(figure 2 (a), $\mathcal L\gg L$).

%%%%%%%%%%%%%%%%%%%%%%%%%%%%%%%%%%%%%%%%%%%%%%%%%%%%%%%%%%
\begin{figure}[!t!b!p]
\centering
\includegraphics [width=1\textwidth] {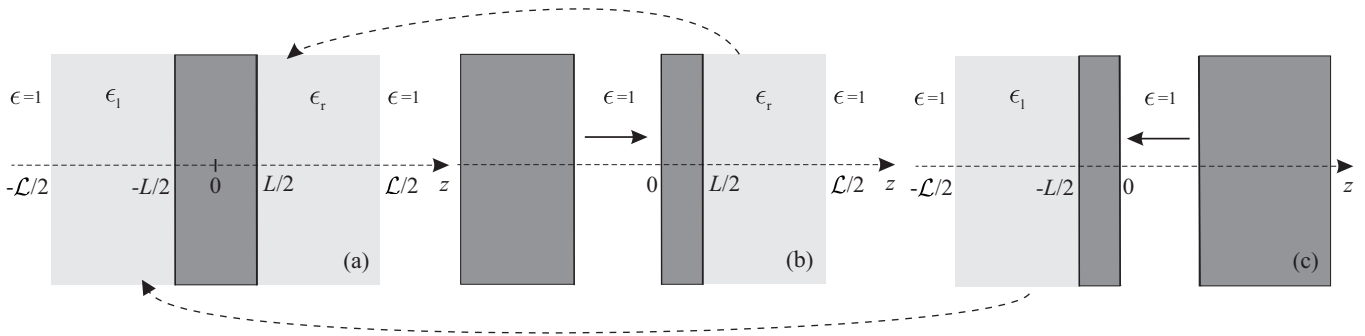}
\caption{(a) -- Scheme of the film in dielectric environment; (b)
and (c) -- split semi-infinite metal samples, which have been in
contact with dielectrics before the splitting. Split parts form a
sandwich in figure 2 (a).} \label{Fig2}
\end{figure}
%%%%%%%%%%%%%%%%%%%%%%%%%%%%%%%%%%%%%%%%%%%%%%%%%

Principal identities for the film can be obtained within a model of
a rectangular well for conduction electrons. To perform a
preliminary analysis, we suppose that the bottom of the potential
well is flat and we count energies starting from its value. Final
expression for the kinetic energies of conduction electrons depends
only on energy differences; therefore, energies counting in such a
way is allowed.

We study a film of thickness $L$ comparable in magnitude to the
Fermi wavelength $\bar{\lambda}_{\rm F}=2\pi\bar{k}_{\rm F}$ of an
electron in 3D metal. The longitudinal sizes of the sample are
assumed to be considerably larger than the film thickness ($L\ll
L_{x},\, L_{y}$), which leads to the pronounced quantization of the
transverse component of the electron momentum. The three-dimensional
Schrodinger equation for a quantum box can be separated into
one-dimensional equations.

The eigenenergies are given by
\begin{equation}
\varepsilon_{ik_{\|}}=\varepsilon_{i}+\frac{k_{\|}^{2}}{2},\,\,k_{\|}^{2}=k_{x}^2+k_{y}^2,
\label{eepp}
\end{equation}
where $\varepsilon_{i}$ is the eigenvalue of the $i$-th
perpendicular state $\psi_{i}(z)$ (hereafter the Hartree atomic
units are used: $\hbar = m = e =1$). The eigenvalue
$\varepsilon_{i}$ is the bottom of the $i$-th subband. For finite
and periodic systems in the z-direction Dirichlet and periodic
boundary conditions are used, respectively. Therefore, possible
allowed electron states ${k_{x}, k_{y}, k_{z}}$ form a system of
parallel planes in the $k-$space, $k_z\equiv k_i$ (see figure 3).

%%%%%%%%%%%%%%%%%%%%%%%%%%%%%%%%%%%%%%%%%%%%%%%%%%%%%%%%%%
\begin{figure}[!t!b!p]
\centering
\includegraphics [width=.3\textwidth] {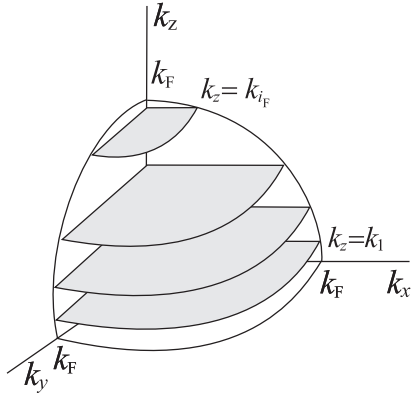}
\caption{Scheme showing occupation of electron states in $k$-space.}
\label{Fig3}
\end{figure}
%%%%%%%%%%%%%%%%%%%%%%%%%%%%%%%%%%%%%%%%%%%%%%%%%

Occupation of electron states starts from the point $\left\{0,0,k_{
1}\right\}$ and follows an increase of radius-vector. As a result,
it turns out that all the occupied states are contained within the
area of $k-$space, confined between the plane $k_{z}=k_{1}$ and
semi-sphere of radius $k_{\rm F}=\sqrt{2\varepsilon_{\rm F}}$.

The number of states $dZ$ in each of the circles, formed by the
intersection of Fermi semi-sphere with planes $k_{z}=k_{i}$ of area
$S=L_{x}L_{y}$, within the interval of wave vectors $(k_\|,
k_\|+dk_\|)$ and taking into account both possible spin projections,
is
$$
dZ(k_\|)=2S\frac{d(\pi k_\|^2)}{(2\pi)^2}.
$$
The maximum value of $k_\|$ in each circle numbered by $i$, is equal
to the circle radius $k_{{\rm F}(i)}=(k_{\rm F}^2-k_i^2)^{1/2}$. In
order to find the number of the occupied states, which coincides
with the number of valence electrons $N$ in the film, one should
integrate $dZ$ over $k_\|$ in each circle, and then sum up
contributions of all the circles:
\begin{equation}
N=\frac{S}{\pi}\sum\limits_{i=1}^{i_{\rm F}}\int\limits_{0}^{k_{{\rm
F}(i)}}{dk_\|\,k_\|}=\frac{S}{2\pi}\left(i_{\rm F}k_{\rm
F}^2-\sum\limits_{i=1}^{i_{\rm F}}k_{i}^2\right). \label{eqN}
\end{equation}

Taking into account an expression for electron kinetic energy
$\frac{1}{2}(k_\|^2+k_{i}^2)$, the total kinetic energy of electron
subsystem equals
\begin{equation}
T_{\rm{s}}=\frac{S}{2\pi}\sum\limits_{i=1}^{i_{\rm
F}}\int\limits_{0}^{k_{{\rm F}(i)}}
dk_\|k_\|\,{\left(k_\|^2+k_{i}^2\right)}
=\frac{S}{4\pi}\sum\limits_{i=1}^{i_{\rm
F}}k_{{\rm{F}}(i)}^2\left(\frac{k_{{\rm{F}}(i)}^2}{2}+k_{i}^2\right),
\label{eqkinen}
\end{equation}
where $i_{\rm F}$ is the number of the last occupied or partially
occupied subband.

In the frame of density-functional theory and stabilized jellium
model (SJ), the total energy of metal sample is represented by the
functional of nonhomogeneous electron concentration $n(\textbf{r})$:
\begin{equation}
E_{\rm SJ}[n(\textbf{r})]=T_{\rm s}+E_{\rm xc}+E_{\rm H}+E_{\rm
ps}+E_{\rm M}, \label{eqen}
\end{equation}
where $T_{\rm s}$ is the (non-interacting) electron kinetic energy,
$E_{\rm xc}$ is the exchange-correlation energy, $E_{\rm H}$ is the
Hartree (electrostatic) energy, $E_{\rm ps}$ is the pseudopotential
(Ashcroft) correction, and $E_{\rm M}$ is the Madelung energy. The
sum of first three terms in expression (\ref{eqen}) corresponds to
the energy of ``ordinary'' jellium, $E_{\rm J}$. The average energy
per valence electron in the bulk of metal is $\bar{\varepsilon}_{\rm
SJ,\,J} =E_{\rm SJ,\,J}[\bar{n}]/N$, where $N$ is a total number of
free electrons of concentration $\bar{n}$, defined by valence and
atomic density.

The positive (ionic) charge distribution can be modeled by the step
function
\begin{equation}
\rho(z)=\bar{n}\theta(L/2-|z|). \label{rhho}
\end{equation}
Solving the Kohn-Sham equations
\begin{equation}
- \frac{1}{2}\nabla^2\psi_i\left(z \right)+ v_{\rm{eff}}
\left[{z,n\left(z\right)}\right]\psi_i\left(z\right)=\varepsilon_i
\psi_i\left(z\right), \label{eqshr_z}
\end{equation}
\begin{equation}
v_{\rm eff}[z,n(z)]=\phi(z)+v_{\rm{xc}}(z)+\left\langle \delta
v\right\rangle_{\rm face}\theta(L/2-|z|) \label{eqv_eff_z}
\end{equation}
together with the Poisson equation
\begin{equation}
\nabla^2\phi(z)=-\frac{4\pi}{\epsilon(z)}\left[n(z)-\rho(z) \right],
\label{eqpois_z}
\end{equation}
with the step function
\begin{equation}
 \epsilon(z)=\left\{
\begin{array}{ll}
   1;           & z< -\mathcal L/2,\, -L/2<z<L/2,\, z>\mathcal L/2,  \\
   \epsilon_{\rm l};  & -\mathcal L/2<z< -L/2, \\
   \epsilon_{\rm r};  &L/2<z< \mathcal L/2, \\
\end{array}
\right. \label{1.99}
\end{equation}
we obtain the single electron wave function and the eigenvalue
$\varepsilon_i$ self-consistently.

We use exchange-correlation potential $v_{\rm
xc}(z)=d[n(z)\varepsilon_{\rm xc}(z)]/dn(z)$ in LDA\cite{Per-Zug}.
The term $\langle\delta v\rangle_{\rm face}$ in (\ref{eqv_eff_z}),
which makes it possible to distinguish different crystal faces,
represents the difference between the potential of the ionic lattice
and the electrostatic potential of the positively charged background
averaged over the Wigner-Seitz cell:
$$
\langle\delta v\rangle_{\rm face}=\langle\delta v\rangle_{\rm
WS}-\left(\frac{\bar{\varepsilon}_{\rm M}}{3}+
\frac{\pi\bar{n}}{6}d^{2}\right),\, \langle\delta v\rangle_{\rm
WS}=- \bar{n} \frac{d\varepsilon_{\rm J}}{d\bar{n}},\label{9}
$$
where $d$ is the distance between the atomic planes parallel to the
surface. The term $\langle\delta v\rangle_{\rm WS}$ describes a
polycrystalline sample \cite{PTS}. In equation (\ref{1.99})
$\epsilon_{\rm l}$ and $\epsilon_{\rm r}$ are dielectric constants
of isolators from the left and right side of the film, respectively.

The electron density profile $n(z)$ is expressed through the wave
functions $\psi_i(z)$
\begin{equation}
n(z)=\frac{1}{2\pi}\sum\limits_{i=1}^{i_{\rm
F}}k_{{\rm{F}}(i)}^2\frac{\left|\psi_i(z)\right|^2}{\int_{-\infty}^{+\infty}dz\left|{\psi_i(z)}
\right|^2}. \label{eqn_z}
\end{equation}
Values of $i_{\rm F}$ and $\varepsilon_{\rm F}$ are determined by
the solution of the equation
\begin{equation}
i_{\rm F}\varepsilon_{\rm F}=\pi L\bar{n}+\sum\limits_{i=1}^{i_{\rm
F}}{\varepsilon_i};\,\, i = 1, 2, \ldots,i_{\rm
F};\,\,\varepsilon_{i}\leq\varepsilon_{\rm F}, \label{cond}
\end{equation}
which follows from the normalization condition (\ref{eqN}) and
definition of the Fermi energy.

In nanofilms, Friedel oscillations are significant throughout the
sample. Therefore, energies are counted from the vacuum level, which
is the energy of the electron in rest in the area $|z|\gg \mathcal
L/2$.  For bound states, energies are negative, including
$\varepsilon _{\rm F}$.

We use iterative procedure (see Appendix A) allowing us to solve
self-consistently the system of equations (\ref{eqshr_z}),
(\ref{eqpois_z}), (\ref{eqn_z}) and to find optimal profiles $n(z)$,
$\phi(z)$, as well as spectrum of one-particle energies. As a
result, work function is defined in the form
\begin{equation}
W=-\varepsilon_{\rm{F}}. \label{eqWF}
\end{equation}

Let us consider a scheme for the surface energy determination (see
figure 2 (b) and (c)) for the film of thickness $L$ in a dielectric
environment.

First, we take a semi-infinite metal (Me$_{\infty}$) covered by a
dielectric ($\epsilon_{\rm r}$). Let us denote the energy of such a
sample as $E\{{\rm Me}_{\infty}|\epsilon_{\rm r}\}$. We now split
the sample and move the parts, as shown in figure 2 (b). As a
result, two new surfaces of the same area $S$ are formed, which are
in a contact with the vacuum ($\epsilon =1$). We denote the energies
of these two parts as $E\{{\rm Me}_{\infty}|1\}$ and $E\{1|{\rm
Me}_{L/2}|\epsilon_{\rm r}\}$, while the irreversible work $A$,
which is needed to form them, as
\begin{equation}
E\{{\rm Me}_{\infty}|1\}+E\{1|{\rm Me}_{L/2}|\epsilon_{\rm
r}\}-E\{{\rm Me}_{\infty}|\epsilon_{\rm r}\}. \label{A1}
\end{equation}
Let us stress that, as a result of these manipulations, the
``fabricated'' sandwich represents a film of thickness $L/2\geq d$
on the dielectric substrate in vacuum (air).

Similar manipulations with another sample (figure 2 (c)) require a
work
\begin{equation}
E\{1|{\rm Me}_{\infty}\}+E\{\epsilon_{\rm l}|{\rm
Me}_{L/2}|1\}-E\{\epsilon_{\rm l}|{\rm Me}_{\infty}\}. \label{A2}
\end{equation}
It is convenient to represent the total energy as a sum of bulk and
surface contributions
$$
E=E^{\rm b}+E^{\rm s}.
$$
Then, bulk components $E^{\rm b}$ do compensate in the expressions
(\ref{A1}) and (\ref{A2}). In each case considered above, the
specific surface energy  $\gamma$ equals $A/2S$.

The work needed to ``create'' a film on a dielectric is
\begin{equation}
A \{\epsilon|{\rm Me}_{L/2}|1\}= \frac{1}{2}\Big [E^{\rm
s}\{\epsilon|{\rm Me}_{L/2}|1\}+E^{\rm s}\{1|{\rm
Me}_{\infty}\}-E^{\rm s}\{\epsilon|{\rm Me}_{\infty}\}\Big].
\label{gamma}
\end{equation}

Now, we join two sandwiches by their free surfaces. We then obtain a
film shown in figure 2 (a). The work to create it can be represented
as the energy of \emph{adhesion} of such two pieces with the minus
sign
\begin{equation}
A\{\epsilon_{\rm l}|{\rm Me}_{L}|\epsilon_{\rm
r}\}=\frac{1}{2}\Big[E^{\rm s}\{\epsilon_{\rm l}|{\rm
Me}_{L}|\epsilon_{\rm r}\}- E^{\rm s}\{\epsilon_{\rm l}|{\rm
Me}_{L/2}|1\} - E^{\rm s}\{1|{\rm Me}_{L/2}|\epsilon_{\rm r}\}\Big].
\label{A3}
\end{equation}
Electron density profiles and potentials for each of the
contributions in the expressions (\ref{gamma}) and (\ref{A3}) are
different, so that they must be calculated self-consistently and
separately.

As similar to the definition for the semi-infinite metal
\cite{LK,Bab-2008}, $E^{\rm s}$ for the film is determined by the
difference between the total film energy (\ref{eqen}) and the energy
of homogeneous metal (stabilized jellium)  of the same volume:
\begin{equation}
E^{\rm s}\{\epsilon_{\rm l}|{\rm Me}_{L}|\epsilon_{\rm r}\}=E_{\rm
SJ}(L)-SL\bar{n}\bar{\varepsilon}_{\rm SJ} = 2S\left\{\gamma_{\rm
J}+\langle \delta v\rangle_{\rm face}
\int\limits_{-L/2}^{L/2}dz\,[n(z)-\bar{n}]\right\}. \label{eqsurfen}
\end{equation}

By using equation (\ref{eqkinen}), quantum-mechanical definition of
an energy
$$
k_{i}^2=-\int\limits_{-\infty}^{\infty}dz\,{\psi_i(z)\nabla^2\psi_i(z)},
$$
as well as the definition given by equation (\ref{eqsurfen}), we
obtain an expression for the first component of $\gamma_{\rm J}$:
\begin{equation}
\gamma_{\rm{s}}=\frac{1}{8\pi}\sum\limits_{i=1}^{i_{\rm
F}}k_{{\rm{F}}(i)}^2\left(\frac{1}{2}k_{{\rm{F}}(i)}^2
-\int\limits_{-\infty}^{\infty}dz\,{\psi_i(z)\nabla^2\psi_i(z)}\right)
-\frac{1}{2}L\bar{n}\bar{t}_{\rm s}, \label{eqkineten}
\end{equation}
where $\bar{t}_{\rm s}=3\bar{k}_{\rm F}^{2}/10$ is the kinetic
energy per 1 electron for bulk. The remaining components are
\begin{equation}
\gamma_{\rm{xc}}=\frac{1}{2}\int\limits_{-\infty}^\infty
dz\,n(z)\varepsilon _{\rm{xc}}[n(z)] -\frac{1}{2}
L\bar{n}\varepsilon_{\rm{xc}}(\bar{n}); \label{eqxcen}
\end{equation}
\begin{equation}
\gamma_{\rm{H}}=\frac{1}{4}\int\limits_{-\infty}^\infty
dz\,\phi(z)\left[n(z)-\rho(z)\right]. \label{eqhartryen}
\end{equation}
For asymmetric sandwiches,  $\{\epsilon_{\rm l}|{\rm
Me}_{L}|\epsilon_{\rm r}\}$, due to the formal division on the
doubled area, the surface energy is calculated ``in average''. This
is the consequence of the definition  of $\gamma$ through the
integral of tangential component of pressure tensor over $z$ from
$-\infty$ to $+\infty$. The pressure tensor contains the
nonelectrostatic part and the Maxwell stress tensor (cf e.g.
\cite{Pg-2006}).

\section{Results and discussion}

We perform calculations for both polycrystalline and crystalline
films made of Na, Al and Pb, with electron concentration
$\bar{n}=3/4\pi r_{s}^{3}$ with corresponding electron parameter
$r_{s}=$ 3.99, 2.07 and 2.30 $a_{0}$.

The minimal thickness of ``crystalline'' sandwiches should be not
less than $2d$. It must be equal to $4d$ for $\{\epsilon_{\rm
l}|{\rm Me}|\epsilon_{\rm r}\}$, only in the case Eq. (\ref{A3}) is
used. d is comparable to $\frac{1}{2}\bar{\lambda}_{\rm F}$
($\bar{\lambda}_{\rm F}=13.06$, 6.78 and 7.53 $a_{0}$ for Na, Al and
Pb, respectively).

%%%%%%%%%%%%%%%%%%%%%%%%%%%%%%%%%%%%%%%%%%%%%%%%%%%%%%%%%%
\begin{figure}
\centering
\includegraphics [width=1\textwidth] {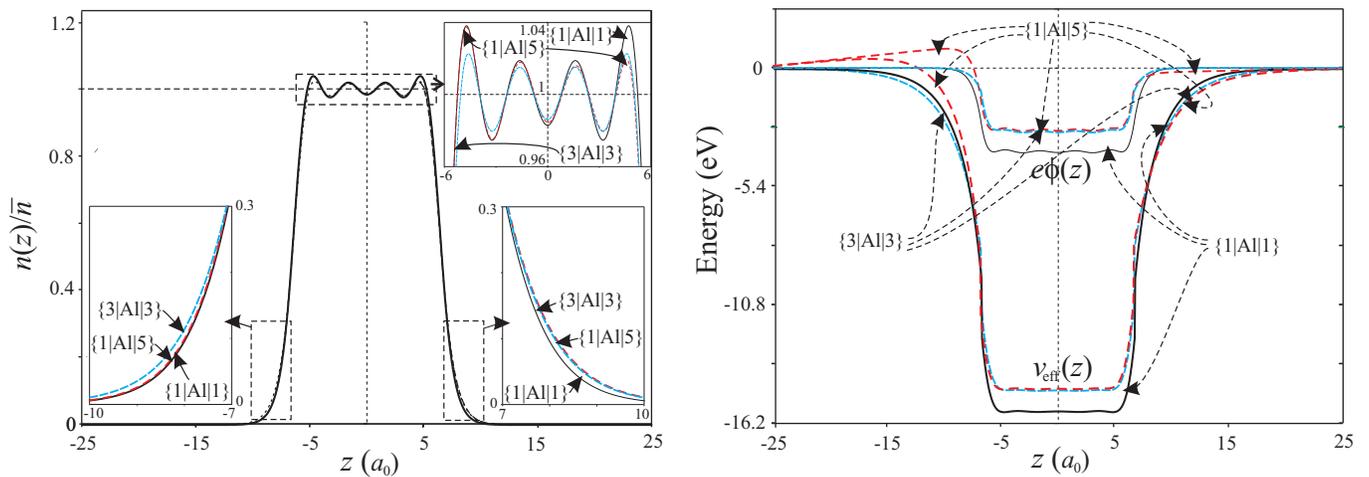}
\caption{The results of self-consistent calculations of the profiles
$n(z)$ of the one-electron effective potential $v_{\rm eff}(z)$, and
the electrostatic potential $\phi(z)$ for sandwiches: $\{1|{\rm
Al}|1\}$, $\{1|{\rm Al}|5\}$ and
  $\{3|{\rm Al}|3\}$ with $L=2\bar{\lambda}_{\rm F}$.}
\label{Fig4}
\end{figure}
%%%%%%%%%%%%%%%%%%%%%%%%%%%%%%%%%%%%%%%%%%%%%%%%%%%%%%%%%%

In view of the multimolecular thicknesses of dielectric coatings  on
the metal film surfaces and rapid fall of the electron distribution
outside of a film (approximately at a distance of 10 -- 15 $a_{0}$),
we formally neglected the effect of a thickness of the coatings,
whose minimum thicknesses must be much greater than that of a
monatomic (molecular) layer of a dielectric. For comparison, we
recall that the diameter of atom Si equals 5 $a_{0}$. If the
thicknesses of the dielectrics on the surfaces of the metal film are
foreseen to be less than the free path of an electron (tens of
angstrom) injected into these dielectrics, then the calculated work
function is qualitatively appropriate for the
vacuum-dielectric-metal slab-dielectric-vacuum.

For symmetric sandwich the effect of a dielectric coating on the
surfaces is reduced to the ``elongation'' of the electron
distribution tail and the effective potential beyond the surface of
a metal (polycrystalline films $\{1|{\rm Al}|1\}$ and $\{3|{\rm
Al}|3\}$ on Fig. 4). The calculations were performed for
$\epsilon=1,\ldots,\,12$. Inside the film one can see the Friedel
oscillations of electron density with peaks near geometrical
boundaries. The period of oscillations is close to
$\frac{1}{2}\bar{\lambda}_{\rm F}$ and only weakly depends on the
presence of dielectric coatings. The situation is similar for Na and
Pb films.

At the boundaries between the metal film and the coatings, there are
jumps in the derivative of the electrostatic potential $\phi'(z)$,
which disappear, provided the dielectric constants of the coatings
are equal to 1. These jumps are due to the boundary conditions
(\ref{Boundary}) at $z=\pm L/2$. The jumps are also reflected on
$v_{\rm eff}(z)$ profile, since $\phi(z)$  is one of its components.
In addition, at the borders, there are another jumps of not only the
derivative $v_{\rm eff}' (z)$, but also of $v_{\rm eff}(z)$ profile
itself for any values of $\epsilon$, including $\varepsilon=1$. Such
jumps have another origin compared to the first ones. This fact is
linked to some features of the model  \cite{PTS}, namely to the
presence of the effective potential component $\left\langle \delta
v\right\rangle _{\rm face} \theta(L/2-|z|)$. These nonphysical jumps
should not be taken into account in the estimation of the effective
force
$$
{\rm {\textbf{F}_{\rm eff}}}(z)\equiv -\nabla v_{\rm eff}(z).
$$
It is seen from figure 4 that force orientations are opposite at
both sides of the film, so that the film in whole must be stressed.
The existence of the force should lead to the increase of spacings
between some lattice planes $d$, while spacings between other planes
must become narrower.

The depth of the potential well $|\bar{v}_{\rm eff}|$, in which the
electrons are located in metal film, decreases ``in average'' with
increasing $\epsilon$ and, as a result, the electron work function
also decreases (see figure 6).

Film spectra $\{1|{\rm Al}_{L}|1\}$ are presented in figure 5. For
comparison, in the same figure, we also provide the results obtained
within the electrons-in-a-box model with the well depth
$U_{0}=-(W_{0}+\bar{\varepsilon}_{\rm F})<0$.

%%%%%%%%%%%%%%%%%%%%%%%%%%%%%%%%%%%%%%%%%%%%%%%%%%%%%%%%%%
\begin{figure}
\centering
\includegraphics [width=.5\textwidth] {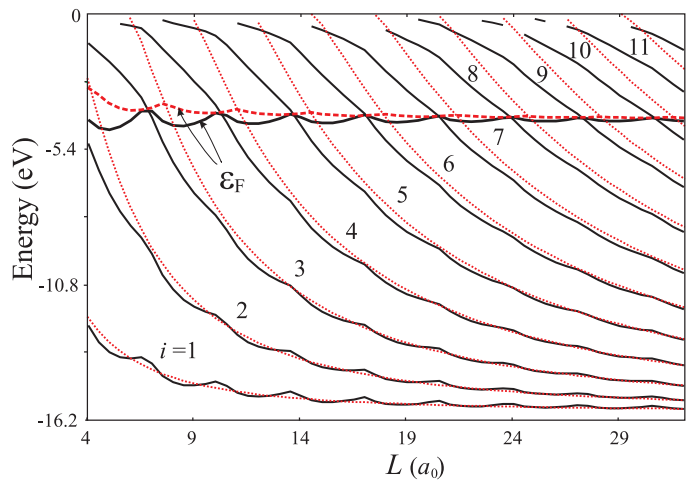}
\caption{Results of calculation for the energy spectrum (subbands)
and Fermi energy $\varepsilon_{\rm F}(L)$ of the film $\{1|{\rm
Al}|1\}$ by the self-consisting method (solid lines) and in
rectangular-box model (dashed lines).} \label{Fig5}
\end{figure}
%%%%%%%%%%%%%%%%%%%%%%%%%%%%%%%%%%%%%%%%%%%%%%%%%%%%%%%%%%

It is seen from figure 5 that the dependence of the eigenstate
energies  on the film thickness, within the SJ model, is oscillating
and decreasing. For subbands with large numbers $i=10,11$, there are
gaps due to the algorithm instability in the vicinity of the vacuum
level. Within the rectangular-box model, this dependence is only
decreasing. Due to smoother edges of the self-consistent well, it
contains more subbands compared to the model of a rectangular box.
Difference in subbands numbers significantly affects calculated
dielectric function and optical conductivity of the nanofilm
\cite{KuPo}.

Within the rectangular-box model, in contrast to the SJ model,
$\varepsilon_{\rm F}(L)$ is always located above one for 3D metal.
Amplitudes of oscillations decrease as $L$ increases. Within both
models, maximum Fermi energies (minimum work functions (\ref{eqWF}))
correspond to the points, in which curves of eigenenergies intersect
Fermi energies. Within the SJ model, in contrast to the
rectangular-box model, minimum Fermi energies correspond to the
points, in which Fermi energy is located between two nearest
eigenenergies (magic film thicknesses similar to magic numbers in
clusters).

Asymmetric sandwiches $\{\epsilon_{\rm l}|{\rm Me}|\epsilon_{\rm
r}\}$ and $\{1|{\rm Me}|\epsilon\}$, which contacts the air or
vacuum, are of particular interest from the viewpoint of
experimental investigation due to the perspective of their use in
technological applications. For instance, these are films on Si
($\epsilon = 12$) or Ge ($\epsilon = 10$) substrates with different
concentration of impurities and crystallographic orientations (see,
e.g. \cite{TAAAA}).

Let us consider both electron density and potential profiles for the
polycrystalline film $\{1|{\rm Al}|5\}$. Presence of a dielectric at
the right side of the film leads to the asymmetry of electron
distribution (see the insets in figure 4), so that there appears a
hump in both the electrostatic and effective potential at the left
side above the vacuum level. This should result, for example, in the
anisotropy of a field emission along the $z-$axis. It is worth
mentioning that bottoms of wells for sandwiches $\{1|{\rm Al}|5\}$
and $\{3|{\rm Al}|3\}$ are essentially the same, some difference
appears only in ``tails'' of potential profiles.

It is of interest to compare heights of humps at $L=10, 12, 13.5$
and 20, 22, 23.5 $a_{0}$. These thicknesses correspond to the
minimum, maximum, minimum of the dependence $W(L)$ for $\{1|{\rm
Al}|5\}$. It turns out that, with the increase of $L$, the hump
height weakly oscillates and decays similarly to the work function,
but maxima of the hump height corresponds to minima of the $W(L)$.
For the values of $L$, as given above, these heights are 0.176,
0.148, 0.170 and 0.158, 0.139, 0.156 eV, respectively.

In order to analyze such a behavior of potential profiles, it is
necessary to go beyond the isotropic model based on a defined
(\ref{rhho}) distribution of \emph{homogeneous} positively charged
background, i.e. one has to take into account not only the reaction
of the electron subsystem, but also the reaction of the ion
subsystem to the presence of a dielectric. Spacings between the
lattice planes are determined by the balance of forces from the
right and left sides for each plane. A simplest realization of this
idea is to disregard variations of spacings between the lattice
planes and to vary the profile of the ion jellium distribution
(\ref{rhho}). We found that such a procedure leads to a significant
deformation of the well bottom, but does not result to considerable
changes of both the spectrum and hump height.

%%%%%%%%%%%%%%%%%%%%%%%%%%%%%%%%%%%%%%%%%%%%%%%%%%%%%%%%%%
\begin{figure}
\centering
\includegraphics [width=1 \textwidth] {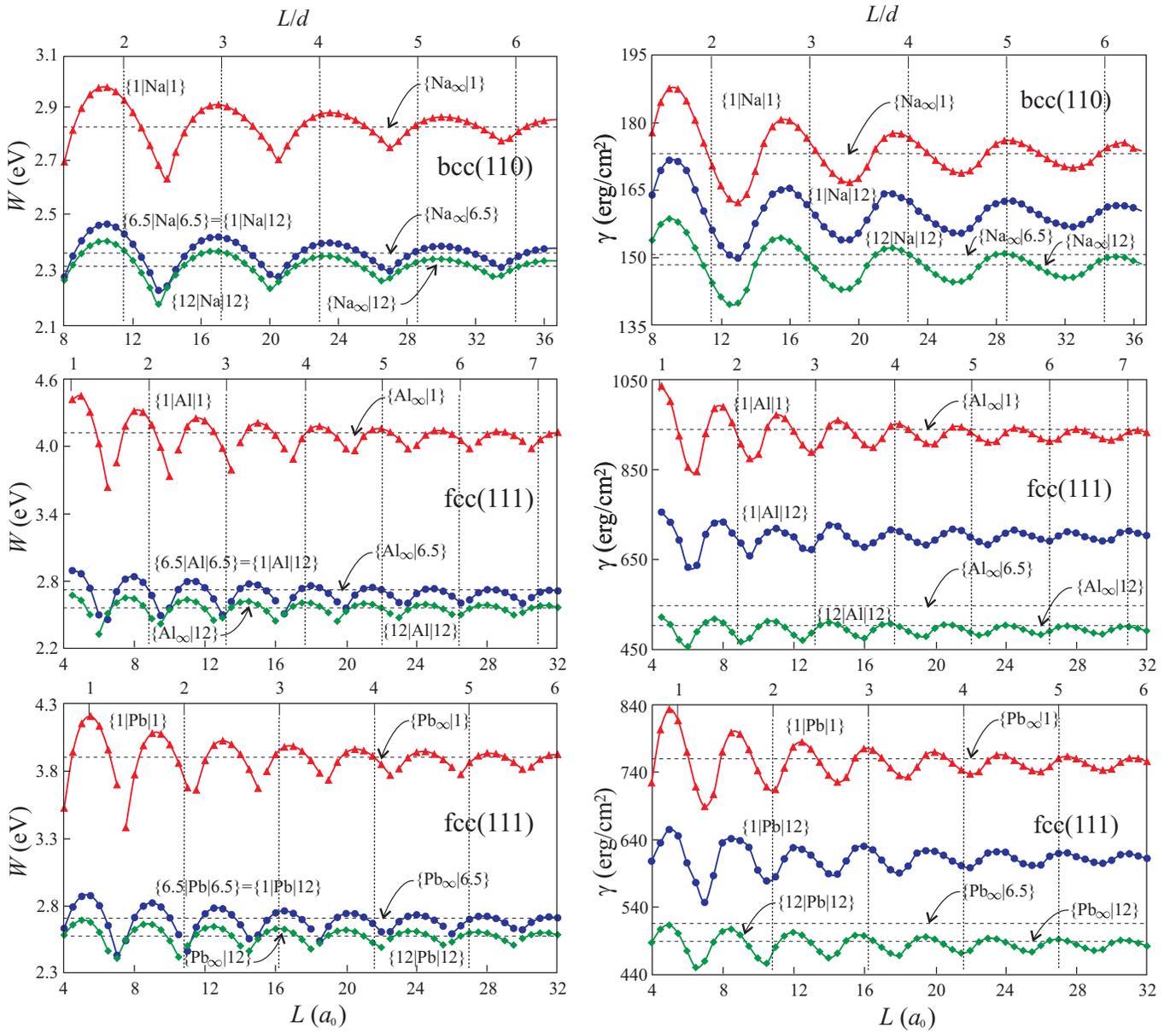}
\caption{Work function and energy per unit of area  for crystalline
sandwiches $\{\epsilon_{\rm l}|{\rm Me}|\epsilon_{\rm r}\}$ and
semi-infinite metal covered by a dielectric $\{{\rm
Me}_{\infty}|\epsilon\}$ (Me $\equiv$ Na, Al, Pb).} \label{Fig6}
\end{figure}
%%%%%%%%%%%%%%%%%%%%%%%%%%%%%%%%%%%%%%%%%%%%%%%%%%%%%%%%%%

Figures 6 show results of our calculations of both the electron work
function and surface energy for crystalline sandwiches using
expression (\ref{eqsurfen}). Horizontal lines correspond to
semiinfinite samples. In contrast to the surface energy, size
dependences $W(L)$ have deep and pronounced minima. It is easier to
analyze them using a simple model  \cite{PKV}. Amplitudes of largest
work function ``oscillations'' are smaller than  0.5 eV. By
considering dependencies for different metals, it is easy to see
that all the differences are due to values of $r_{s}$. For the Al,
which has the smallest $r_{s}$, work function oscillations are
maximum, while the period is minimum. Positions of both maxima and
minima depend weakly on $\epsilon$ of a dielectric and slightly
shift towards smaller $L$ with the increase of $\epsilon$.

In contrast to the work function, surface energy oscillations can be
approximated by analytical dependences
$$
E^{\rm s}\{\epsilon |{\rm Me}_{L}|\epsilon\}= E^{\rm s}\{{\rm
Me}_{\infty} |\epsilon\}+{\mathcal A} \frac{\sin \left(2\bar{k}_{\rm
F}L + \varphi\right)}{L},
$$
with parameters ${\mathcal A}$ and $\varphi$. Maxima of function
$W(L)$, $\gamma (L)$ correspond to ``magic'' film thicknesses, which
are defined by maximum occupation of a given subband.

The unexpectable result of self-consistent calculations is a
coincidence of dependencies  $W(L)$ for sandwiches $\{1|{\rm
Me}|12\}$ and $\{6.5|{\rm Me}|6.5\}$. Computations for $\{1|{\rm
Me}|5\}$ and $\{3|{\rm Me}|3\}$ give a similar result. This means
that the electron work function for asymmetric sandwiches
$\{\epsilon_{\rm l}|{\rm Me}|\epsilon_{\rm r}\}$ coincides with high
accuracy with the work function for symmetric sandwiches
$\{\langle\varepsilon \rangle|{\rm Me}|\langle\epsilon \rangle\}$
with the averaged value $\langle\epsilon \rangle
=\frac{1}{2}(\epsilon_{\rm l}+\epsilon_{\rm r})$.

Work function has both the bulk and surface contributions. Because
bulk metal contributions $W(L)$ for sandwiches $\{1|{\rm
Me}_{L}|12\}$ and $\{6.5|{\rm Me}_{L}|6.5\}$ are the same by
definition, also the same are contributions of dipole surface
barriers. We here imply the total contribution of both sides of a
sandwich, since the work function is an ``\emph{isotropic}''
characteristics\cite{25a}. A coincidence of work functions is most
likely a geometric effect. This feature will be addressed elsewhere.

For surface energies, such a coincidence does not exist. It is not
difficult to perform calculations according to formulas
(\ref{gamma}) and (\ref{A3}), if $\gamma$ are known.

\section{CONCLUSIONS}

We proposed a method for the self-consistent calculations of
spectra, electron work function, and surface energy of metal films
placed into passive dielectrics. As typical examples, we considered
Na, Al, and Pb films.

The effective force acting on the film from the outside is due to
the inhomogeneous electron distribution. This force should lead to
film stressing in a transverse direction. The effect of the
stressing generally becomes more significant with the increase of
the film thickness.

In contrast to the surface energy, size dependencies of work
function have deep and strongly pronounced minima. The smaller
$r_{s}$ the more difficult the problem of numerical analysis of size
dependencies in the vicinities of these minima.

With the increase of film thickness up to few $\bar{\lambda}_{\rm
F}$, size variations of both the work function and surface energy
occur near their average values (for symmetric sandwiches, these
values correspond to 3D metals and do not contain significant
monotonous size contributions). Dielectric environment generally
leads to the decrease of electron work function and surface energy.

We also considered asymmetric metal-dielectric sandwiches
characterized by different dielectrics at both sides of the film.
One of the examples of such systems is a film on the dielectric
substrate. We found that the presence of a dielectric from one side
of the film leads to such a ``deformation'' of electron distribution
that there appears a ``hump'' above the vacuum level both in the
electrostatic and effective potentials. The asymmetry of potential
profile of this kind should lead to an anisotropy of the field
emission. In addition to size dependencies, the shift of the work
function is generally determined by the average dielectric constants
of environments.

Obtained wave electron functions for metal-dielectric sandwiches
allow us to calculate the matrix elements of optical transitions,
conductivity tensor, and coefficient of optical
absorption\cite{KuPo}.

\acknowledgments

We thank  W. V. Pogosov for reading the manuscript.

\begin{appendix}

\section{Self-consistency procedure}

The initial approximation $n(z)$ is chosen for solving the Kohn-Sham
equations in the form of a one-parametric trial function
$n^{(0)}(z)=\bar {n}f(z)$, where
$$
f(z)=\left\{{\begin{array}{ll}
 -\frac{1}{2}{e^{(z-L/2)/\lambda}+\frac{1}{2}e^{(z+L/2)/\lambda}}, & z<-L/2, \\
 1-\frac{1}{2}{e^{(z-L/2)/\lambda}-\frac{1}{2}e^{-(z+L/2)/\lambda}}, & |z|<L/2, \\
 -\frac{1}{2}{e^{-(z+L/2)/\lambda}+\frac{1}{2}e^{-(z-L/2)/\lambda}}, & z>L/2. \\
\end{array}}\right.
$$
$\lambda$ is the variational parameter, which is found through the
minimization of surface energy. Solution by a direct variational
method is an independent problem, which is not addressed in this
paper (for simple metals $\lambda$ is closed to $1\, a_{0}$). As a
result of integration of equation (\ref{eqpois_z}), within the
initial approximation, we obtain
$\phi^{(0)}(z)=-4\pi\bar{n}\lambda^2 f(z)$.

Each wave function $\psi(z)$ is constructed as
$$\psi(z)=\left\{{\begin{array}{ll}
 \psi_{\rm left}(z), & z<z_0, \\
 \psi_{\rm right}(z), & z>z_0, \\
\end{array}}\right.$$
under the condition of continuity of functions $\psi_{\rm
left}(z_0)=\psi_{\rm right}(z_0)$, as well as of their derivatives
$\psi'_{\rm left}(z_0)=\psi'_{\rm right}(z_0)$. $z_0$ is an
arbitrary point in the interval $z\in[-L/2;+L/2]$, while $\psi_{\rm
left}(z)$ and $\psi_{\rm right}(z)$ are functions, which are found
by a numerical solution of Eq. (\ref{eqshr_z}) by the Numerov`s
method from $z=z_{-}$ to $z=z_0$ and from $z=z_{+}$ to $z=z_0$,
respectively. It is sufficient to take values
$z_{\mp}=\mp(L+20)\,a_{0}$. In these points, the potential profile
 $v_{\rm{eff}} (z)$ is cut off. The boundary conditions (\ref{eqshr_z})
here are determined by the behavior of the wave function $\psi$
under the barrier from the left ($e^{z\sqrt {|\varepsilon_i|}}$) and
right ($e^{-z\sqrt {|\varepsilon_i|}}$) sides from the slab
($|z|\geq |z_{\mp}|$), respectively. Boundary conditions provide
wave function, as well as its derivative at $z=z_{\mp}$. This
peculiarity of our computations is due to the fact that errors of
the numerical method for the wave function $\psi_{\rm right}(z)$ and
$\psi_{\rm left}(z)$ near the right and left boundaries of the
interval grow, since the round-off errors also increase and lead to
the instability of the algorithm under the motion towards the
exponential damping.

In order to solve the system of equations (\ref{eqshr_z}),
(\ref{eqpois_z}) and (\ref{eqn_z}) self-consistently, with
relatively small number of iteration steps, the Poisson equation
(\ref{eqpois_z}) should be modified, in particular, by introducing a
perturbation \cite{Man1975}.

Equation (\ref{eqpois_z}) is solved by the Lagrange method in the
form
\begin{equation}
{\phi^{(j)}}''-q^2\phi^{(j-1)}=-\frac{4\pi}{\epsilon(z)}\left[n^{(j)}-\rho\right]
- q^2\phi^{(j-1)} \label{eqpois_mod}
\end{equation}
with the boundary conditions
\begin{equation}
 \begin{array}{ll}
    \phi^{(j)}_{\rm out}(z)=\phi^{(j)}_{\rm in}(z),\,\, {\epsilon_{\rm l}\phi^{(j)}_{\rm out}}'(z)={\phi^{(j)}_{\rm
in}}'(z);\,\,\,           & z=-L/2,  \\
    \phi^{(j)}_{\rm in}(z)=\phi^{(j)}_{\rm out}(z),\,\, {\phi^{(j)}_{\rm
in}}'(z)={\epsilon_{\rm r}\phi^{(j)}_{\rm out}}'(z);\,\,\,           & z=L/2,  \\
    {\phi^{(j)}_{\rm out}}(z)=0,\,\,{\phi^{(j)}_{\rm out}}'(z)=0; \,\,\, & z=\mp\infty. \\
   \end{array}
\label{Boundary}
\end{equation}
The term $q^2\phi$ was introduced as a small perturbation;
$\phi_{\rm out}(z)$ and $\phi_{\rm in}(z)$ are potentials outside
and inside the film, respectively. In equation (\ref{eqpois_mod}),
at each step of the iteration $j=1,\,2,\,3,\,...$, electrostatic
potential profile depends not only on the electronic concentration
profile, but also on its own profile at the previous iteration. It
is convenient to take $q$ equal to electron wave number at the Fermi
sphere $\bar{k}_{\rm F}= (3\pi^{2}\bar{n})^{1/3}$ of homogeneous
electron liquid.

The solution of equation (\ref{eqpois_mod}) for $\mathcal L\to
\infty$ has the simple form
\begin{widetext}
\begin{equation}
\begin{array}{l}
\phi^{(j)}(z)= \left\{{\begin{array}{ll}
 \left(\int\limits_{-\infty}^z{{\frac{e^{ - q{z}'}}{2q}f_1 d{z}'
+ A_1 }}\right)e^{qz} + \left(-\int\limits_{-\infty}^z
{{\frac{e^{q{z}'}}{2q}f_1d{z}'+B_1}}\right)e^{-qz}, & z < - L / 2, \\
 \left(\int\limits_{-L/2}^z {{\frac{e^{-q{z}'}}{2q}f_2 d{z}'+
A_2}}\right)e^{qz} + \left(-\int\limits_{-L/2}^z
{{\frac{e^{q{z}'}}{2q}f_2 d{z}'+B_2}}\right)e^{- qz}, & |z|\le L/2, \\
 \left(-\int\limits_z^{\infty}{{\frac{e^{- q{z}'}}{2q}f_3 d{z}'
+A_3}}\right)e^{qz} + \left(\int\limits_z^{\infty}{{\frac{e^{q{z}'}}{2q}f_3 d{z}'+B_3}}\right)e^{- qz},& z>L/2, \\
\end{array}} \right.\\
\end{array} \label{eqphi_z}
\end{equation}
\end{widetext}
where $f_{m}({z}')=-4\pi
[n({z}')-\rho({z}')]D_{m}-q^2\phi^{(j-1)}({z}')$ and $D_{m}=$
$\epsilon_{\rm l}^{-1},\,1,\,\epsilon_{\rm r}^{-1}$ for $m=1,2,3$,
respectively. The choice of values  $B_1=0$ and $A_3=0$ immediately
follows from the condition of finiteness of potentials far away from
the film.

Values of other coefficients $A$ and $B$ are found from the solution
of the system of equations (\ref{Boundary}):
$$
 A_1=\frac{2A_2}{1+\epsilon_{\rm l}}+\frac{1-\epsilon_{\rm l}}{1+\epsilon_{\rm l}}
 \int\limits_{-\infty}^{-L/2}\frac{e^{q({z}'+L)}}{2q}f_1 d{z}'-
 \int\limits_{-\infty}^{-L/2}\frac{e^{-q{z}'}}{2q}f_1 d{z}',
$$
$$
B_3=\frac{2B_2}{1+\epsilon_{\rm r}}-\frac{1}{1+\epsilon_{\rm
r}}\int\limits_{-L/2}^{L/2}\frac{e^{q{z}'}}{q}f_2 d{z}' +
\frac{1-\epsilon_{\rm r}}{1+\epsilon_{\rm
 r}}\int\limits_{L/2}^{\infty}\frac{e^{-q({z}'-L)}}{2q}f_3 d{z}'-
 \int\limits_{L/2}^{\infty}\frac{e^{q{z}'}}{2q}f_3 d{z}'.
$$

Let`s introduce notation
\begin{multline}
J_{(\pm)}=
 Y_{0}\left[Y_{1}2\epsilon_{\rm l}(1\mp\epsilon_{\rm
 r})\int\limits_{-\infty}^{-L/2}d{z}'e^{q{z}'}f_1\right.
 + \left.Y_{2}(1\pm\epsilon_{\rm l})(1+\epsilon_{\rm r})\int\limits_{-L/2}^{L/2}d{z}'e^{-q{z}'}f_2\right.
 +\left.Y_{3}(1\pm\epsilon_{\rm l})(1-\epsilon_{\rm
 r})\int\limits_{-L/2}^{L/2}d{z}'e^{q{z}'}f_2\right.\\
 + \left.Y_{4}2\epsilon_{\rm r}(1\pm\epsilon_{\rm l})\int\limits_{L/2}^{\infty}d{z}'e^{-q{z}'}f_3\right],
\end{multline}
where $Y_{0}=\{2q[(1-\epsilon_{\rm l})(1-\epsilon_{\rm
r})e^{-qL}-(1+\epsilon_{\rm l})(1+\epsilon_{\rm r})e^{qL}]\}^{-1}$.
Then $A_{2}=J_{(+)}$ for $Y_{1,3}=1,\, Y_{2,4}=e^{qL}$ and
$B_{2}=J_{(-)}$ for $Y_{2,4}=1,\, Y_{1}=e^{qL}=Y_{3}^{-1}$.

In the case of the symmetric sandwich  $\epsilon_{\rm
l}=\epsilon_{\rm r}$ the accurateness of calculations is verified by
examination the stationarity conditions $n'(z)=0$ and
${\phi^{(i)}_{\rm in}}'(z)=0$ in the center of the slab ($z=0$).

\end{appendix}

\end{document}